# Logical and information aspects in surface science: friction, capillarity, and superhydrophobicity

Michael Nosonovsky[a]*

[a]*Department of Mechanical Engineering, University of Wisconsin-Milwaukee, Milwaukee, WI 53211-0413 USA*

* nosonovs@uwm.edu, Tel. +1-414- 229-2816

Provide short biographical notes on all contributors here if the journal requires them.

# Logical and information aspects in surface science: friction, capillarity, and superhydrophobicity

Logical and information aspects of friction and wetting (including the adhesion, capillarity, and superhydrophobicity) are discussed. Friction involves paradoxes, such as the Painlevé paradoxes of non-existence or non-uniqueness of solutions in mechanical systems of rigid bodies with dry friction. These paradoxes can be treated by introducing ternary logic with the three basic states: rest-motion-undefined. When elastic deformation is introduced, the paradoxical solutions correspond to frictional instabilities leading to rest-motion-unstable as three states of a system. The dynamic evolution of a frictional interface towards a limit cycle can be viewed as a process of erasing the information about the interface due to the instabilities. Furthermore, while friction force is universal, it is not treated as a fundamental force and can be considered as an epiphenomenon of various synergetic mechanisms. This further relates friction to other surface effects, including the capillarity, with its binary logic of wetting states and a possibility of droplet computation for lab-on-a-chip microfluidic reactors. We discuss the logical foundation of biomimetic superhydrophobic surface design and how it is different from the conventional design. Both friction and wetting can be used for novel unconventional logical and computational devices.

## 1. Introduction

Dry sliding friction is a fundamental physical phenomenon which occurs almost universally for all classes of materials (metals, polymers, ceramics, composites, etc.) and material combinations. Friction is found for a large range of loadings, from nanonewtons (in nanotribology) to billions of tons (in seismology) and for a wide range of other conditions. On the other hand, the friction force is not considered a fundamental physical force, and it is usually introduced in an ad hoc manner, rather than deduced from the first principles of physics. There are many seemingly unrelated mechanisms leading to friction. These mechanisms include adhesion, elastic and plastic deformation, brittle fracture, rupture crack propagation, etc.

Due to this ambiguous status of friction in physics – universal on the one hand and unrelated to the fundamental principles of physics on the other hand – combining friction with the rest of mechanics and physics involves a number of interesting logical paradoxes. Furthermore, friction and wetting can be used for unconventional logical and computational devices with parallel and distributed computation combined with a chemical reaction. In this paper we discuss logical and information aspects of friction and a broader range of surface effects, such as the adhesion and superhyrophobicity in biomimetic surfaces [1-3].

## 2. Friction from the logical and information point of view

Friction has implications related to logic and information. This includes frictional logical paradoxes of non-existence and non-uniqueness, friction-induced self-organization with is naturally characterized by information entropy, and the status of friction as a universal phenomenon which, however, cannot be reduced to a fundamental force.

### *2.1. Dynamic Friction and Logical Paradoxes*

The best known family of frictional paradoxes are the Painlevé paradoxes involving rigid (non-deformable) bodies and Coulomb friction, named after Paul Painlevé (1863-1933), a French mathematician and politician (he became a Prime-Minister of France in 1917). To solve the equations of statics, an assumption should be made about the direction of the friction force. However, after the solution is obtained, it may turn out that the assumed direction of the friction force contradicts the direction of the velocities in the system, therefore resulting in a paradox [1].

In the system shown in Fig. 1, two sliders both having the mass *m* are connected by a link with a constant length *l* forming the angle of $\varphi$ with the sliding surface. The upper slider is frictionless while the lower slider is frictional with the coefficient of friction µ. An external force *P* is applied to the upper slider. The motion of such a system is governed by the equation $2m\ddot{x} = P - \mu|R|\text{sign}(\dot{x})$,, where *R* is the normal force acting at the first slider ($R\cos\varphi$ is the compression force in the link). From the balance of forces acting on the second slider, $m\ddot{x} = P + R/\tan(\varphi)$.

To find the unknown $\ddot{x}$ and *R*, we should assume a value of $\text{sign}(\dot{x})$. However, if $\mu \tan \varphi > 2$ then two solutions exists for $\dot{x} > 0$ satisfying $m\ddot{x} = P(1 \pm \mu \tan \varphi)/(2 \pm \mu\tan \varphi)$, while no solution exists for $\dot{x} < 0$.

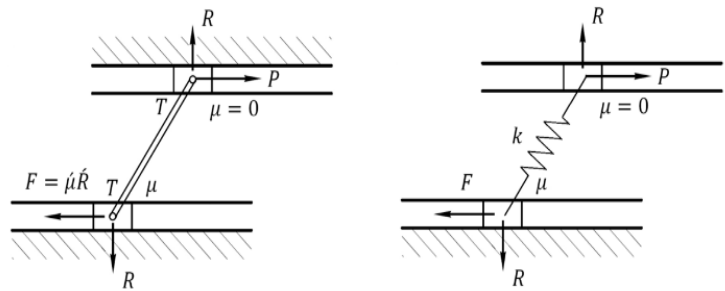

Figure 1. Setup for the Painlevé paradox: two sliders (one frictional and the other is frictionless) connected by a rigid bar; the paradox is resolved if the bar is assumed to be elastically deformable with the compliance *k*.

The Painlevé paradox indicates that the Coulomb friction is not always logically compatible with the rest of the equations of mechanics. In a formal manner this can be presented in the following way. Consider a predicate P(α, β…) defined over a system of mechanical equations with a set of parameters α, β,... (e.g., characterizing system's geometry or properties such as friction) and corresponding initial conditions with P= true, when the equations have a unique solution, and P(α, β…)=false otherwise. For a mechanical system without friction for any values of parameters and for any initial conditions there is always a solution

$$\forall (\alpha, \beta, \ldots) \text{P}(\alpha, \beta, \ldots) = T \qquad (1)$$

However, for a system with the Coulomb friction, such as in Fig. 1, the solution exists when and only when parameters are within the specified range

$$(\mu \tan \varphi \leq 2) \Leftrightarrow \text{P}(\mu, \varphi). \qquad (2)$$

Furthermore, a predicate M(α, β…) can be introduced with meaning "the system is in motion." In the context of friction, these two states are typically called the "slip" and "stick" states. Typically, M(α, β…) is either true or false

$$\forall (\alpha, \beta, \ldots) \text{M}(\alpha, \beta, \ldots) \oplus \neg \text{M}(\alpha, \beta, \ldots), \qquad (3)$$

where ⊕ stands for the exclusive disjunction ("exclusive or") operator and ¬ is the negation. When the paradox exists

$$(\mu \tan \varphi \leq 2) \Leftrightarrow \text{M}(\mu, \varphi) \oplus \neg \text{M}(\mu, \varphi) \qquad (4)$$

is equivalent to

$$(\mu \tan \varphi > 2) \Leftrightarrow \big(\text{M}(\mu, \varphi) \wedge \neg \text{M}(\mu, \varphi)\big). \qquad (5)$$

Therefore, the paradox essentially violates the classical law of excluded middle (*tertium non datur*) of the binary Aristotelean logic. However, it may be valid in a more complex ternary logic, in which some mechanical systems may be neither in motion nor at rest, and a third value, M=undefined, is possible, so that.

$$(\mu \tan \varphi > 2) \Leftrightarrow \text{M}(\mu, \varphi) = \text{undefined} \qquad (6)$$

While this brings the memory of the classical "Zeno's arrow paradox," it should be shown whether the ternary logic can be productive for any problems of classical mechanics.

There is a significant literature devoted to the nature of the Painlevé paradox and various approaches to resolve it [4-10]. Most approaches either modify the law of friction by suggesting a non-Coulombian friction law, e.g., velocity dependent [4], or consider deformable bodies instead of the rigid ones [5-7]. Moreau used classical concepts of Convex Analysis and measure differential inclusions to develop a numerical scheme for paradoxical cases [9]. Stewart [9] applied "sweeping processes" and the measure differential inclusions along with the variational inequality approaches [9]. Génot and Brogliato used the close relationship of the paradoxes to the so-called linear complementary problem [10]. From a more physical, rather than mathematical, point of view, there is an interesting relationship between the Painlevé paradoxes and the friction-induced instabilities [1].

If an elastically deformable link is considered instead of the rigid link with the compliance $k$ (a non-negative parameter defined as the inverse of the stiffness or of the elastic modulus), the sliding system obtains an additional degree of freedom. In that case, the paradox corresponds to the unstable solution with the reaction force growing until the value of φ decreases so that the paradox condition $\mu \tan(\varphi) > 2$ will not be satisfied anymore. Thus the static paradox of a non-existent solution, when studied in dynamics, corresponds to an unstable solution.

In a formal manner this can be formulated as a new (binary) predicate S(α, β…) meaning "the motion/equilibrium is stable" is introduced

$$(k \neq 0) \wedge (\mu \tan \varphi \leq 2) \Longrightarrow S(\mu, \varphi, k) \tag{7}$$

$$(k \neq 0) \wedge (\mu \tan \varphi > 2) \Longrightarrow \neg S(\mu, \varphi, k) \tag{8}$$

$$(k = 0) \wedge (\mu \tan \varphi \leq 2) \Longrightarrow \big(M(\mu, \varphi, k) \oplus \neg M(\mu, \varphi, k)\big) \tag{9}$$

$$(k = 0) \wedge (\mu \tan \varphi > 2) \Longrightarrow M(\mu, \varphi, k) = \text{undefined} \tag{10}$$

It is concluded from Eqs. (7)-(10) that in the limit of small compliance, $k \to 0$, when the deformable link becomes a rigid one, the instable motion becomes equivalent to the paradox (in the binary logic) or to the undefined state of the predicate M (in the ternary logic):

$$\neg S(\mu, \varphi, k) \Leftrightarrow M(\mu, \varphi, k) = \text{undefined, for } k = 0. \qquad (11)$$

In other words, the ternary logic of the rigid (*k*=0) system with predicate's values corresponding to "the system is at rest / moving / undefined" obtains a new interpretation for compliant ($k \neq 0$) systems: "at rest / moving / unstable."

Note that in general the introduction of the compliance does not resolve the paradox. Adams and co-workers demonstrated that dynamic effects in elastically deformable (compliant) systems lead to new types of frictional paradoxes with the assumed direction of sliding used for the Coulomb friction opposite to the resulting slip velocity [11]. In a strict mathematical sense, the Coulomb friction law is inconsistent not only with the rigid body dynamics, but also with the dynamics of the elastically deformable bodies. Therefore, a modified law should be introduced, such as the rate-and-state friction law widely used in geo-mechanics and theories of dynamic friction [1].

Despite that, considering the *unstable motion* as a third possibility besides the *rest* and *stable motion* is very productive. Investigating the unstable motion introduces a new and diverse class of mechanical phenomena, which were rarely investigated by mechanicians until the end of the 20[th] century, including those leading to the self-organization and hierarchical structures. In the next section, friction-induced instabilities and their applications will be discussed.

*2.2. Frictional instabilities*

Friction is usually thought of as a stabilizing factor; however, sometimes friction leads to the dynamic instability of sliding. The stability of sliding of two pure elastic half-spaces with a constant coefficient of friction between them is a relatively simple mathematical problem, which could have been solved already in the 19$^{th}$ century (Fig. 2). However, it was not thoroughly studied until the 1990s when Adams [12] discovered that, for a broad range of material parameters, the motion is dynamically unstable. An elastic wave can propagate along the interface between two slightly dissimilar (in terms of their elastic properties) elastic bodies with no friction between them. This wave is confined to the interfacial area, because the wave magnitude decreases exponentially with the distance from the interface. This is the so-called generalized Rayleigh wave, which is a generalization of the concept of the elastic surface wave, called the Rayleigh wave.

If a small constant coefficient of friction is introduced, the amplitude of the generalized Rayleigh wave will not remain constant anymore. Instead, the amplitude will grow with time in an exponential manner, making the sliding dynamically unstable. The source of energy for these growing-amplitude waves is in the work done by the external force applied to overcome friction. A similar effect was found for rough surfaces as well [13].

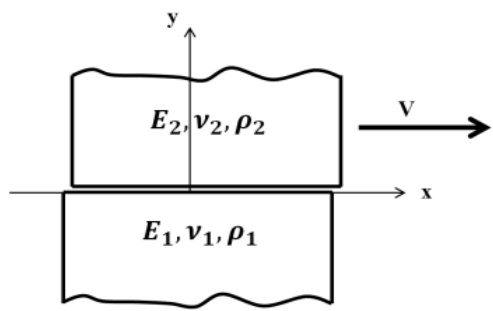

Figure 2. Two elastic half-spaces (characterized by the elastic moduli, $E_1$ and $E_2$, Poison's ratios $\nu_1$ and $\nu_2$ and densities $\rho_1$ and $\rho_2$) slide relative to each other with the velocity *V*. For slightly dissimilar materials (in terms of their elastic properties) an interface elastic wave can propagate, and the wave becomes unstable when friction is introduced.

Instabilities of different types emerge when the coefficient of friction decreases with increasing sliding velocity or when friction is coupled with another effect, such as the thermal expansion of the material. A general thermodynamic stability criterion has been suggested to study any type of instabilities. The criterion states that the second variation of the rate of entropy (*s*) production should be positive, $\delta^2 \dot{s} > 0$, in order for the motion to remain stable [1]. The expression for the entropy rate can include mechanical, thermal, chemical, electric, micro/nanostructural, and other components. Using the stability logical predicate S as introduced in the preceding section, one can define

$$S \equiv (\delta^2 \dot{s} > 0) \tag{12}$$

If only mechanical interactions occur in a frictional system with one degree of freedom, then the rate of entropy is given by the rate of energy dissipation (the product of the sliding velocity *V*, and the friction force µW) divided by temperature, $\dot{s} = \mu W V / T$. If two of these parameters are varied and interrelated, we obtain

$$\delta^2 \dot{s} = \frac{2W}{T} \delta V \delta \mu = 2W \frac{d\mu}{dV} (\delta V)^2 > 0 \text{ or } \frac{d\mu}{dV} > 0 \tag{13}$$

Using the definition in Eq. (12), one should note that the predicate $S(\delta V, \delta \mu)$ depends on the *variations* of the coefficient of friction and sliding velocities, rather than on the values of these parameters, keeping in mind $\delta V \neq 0$, *T* > 0 to avoid the division by zero. One can write then a formal proof

$$\frac{d\mu}{dV} > 0 \Rightarrow \frac{\delta \mu}{\delta V} > 0 \Rightarrow \frac{2W}{T} \delta V \delta \mu > 0 \Rightarrow \delta^2 \dot{s} > 0 \Rightarrow S(\delta V, \delta \mu). \tag{14}$$

In other words, if the coefficient of friction decreases with the sliding velocity, the system is unstable. This is because an increasing sliding velocity causes a decrease in the frictional resistance and further increase of the velocity.

The interactions, which occur when frictional sliding is coupled with another process that contributes to entropy production, such as thermally activated material transfer, chemical reactions, or wear [1], are much more interesting and complex than pure mechanical interactions. An example will be presented in the next section.

The study of an unstable motion has been traditionally a marginal area of mechanics. This was, in part, caused by the binary logic with emphasized the "stable motion" and "rest" states however marginalized the unstable motion. The advances of recent decades in studying non-equilibrium process made it clear that the "unstable motion" is a very important area of research. On the one hand, it can lead to engineering applications, such as the development of new materials. On the other hand, frictional systems can lead to new logical and informational devices, if friction is viewed from the viewpoint of information production, rather than from the view point of energy dissipation. The instabilities are the driving forces of this informational approach.

## *2.3. Frictional evolution and information*

Frictional instabilities can lead to the formation of new structures at the interface, especially when friction is coupled with another process, such as thermally activated material expansion, chemical reaction, wear, mass transfer of one phase in a composite material, electric current, and many others. These self-organized friction-induced patterns or "secondary structures" can include a broad range of phenomena including *in situ* formed tribofilms, patterns of surface topography, and other interfacial patterns like propagating trains of stick and slip zones formed due to dynamic sliding instabilities

[1].

The underlying mathematics of these processes is very similar to what we have seen in the preceding section. Suppose now that the coefficient of friction depends on a microstructure parameter ϕ, such as the thickness of the interface film. The stability condition is now given by

$$\delta^2 \dot{s} = \delta\left(\frac{\mu W V}{T}\right)\delta\phi = \frac{W}{T}\frac{\partial \mu}{\partial \phi}V(\delta\phi)^2 > 0 \qquad (15)$$

In this case, the derivative $\partial\mu/\partial\phi$ is itself a function of ϕ, so there may be a critical value of $\phi_{cr}$ (e.g., critical thickness of a tribofilm), above which the derivative changes its sign and the system destabilizes, e.g., $\partial\mu/\partial\phi < 0$ if $\phi > \phi_{cr}$).

If the stability condition is violated for a certain value of ϕ, then further growth of the film will result in decreasing friction, which will facilitate the further growth of the film. Note that Eq. 15 has the same structure as Eq. 13 if ϕ=V.

Using the definition of Eq. 12, the condition for the growth of the protective tribofilm becomes

$$\phi > \phi_{cr} \Rightarrow \frac{\delta\mu}{\delta\phi} > 0 \Rightarrow \frac{2WV}{T}\delta\phi\delta\mu > 0 \Rightarrow S(\delta V, \delta\mu). \qquad (16)$$

Experimental observations show that friction can indeed lead to the formation of in situ protective tribofilms and more complex structures at the frictional interface. Such reaction occurs in a number of situations when a soft phase is present in a hard matrix, including Al-Sn and Cu-Sn-based alloys [1]. Other examples are the lead-bronzes interface, martensite surface layers in steel, and carbon reduction during the friction of copper. The in-situ formed tribofilm can have protective properties and reduce friction and wear; therefore, it is desirable to find the conditions under which such film would grow. Proper understanding and control of these processes allows developing of new tribomaterials.

Besides the growth of tribofilms, there is another frictional evolutionary process. When frictional sliding starts, at first the coefficient of friction usually higher than in the steady state regime. The initial non-stationary regime is usually called *run-in* [1]. During this regime the surfaces tend to adjust to each other, for example, by changing roughness parameters due to an extensive deformation and fracture of asperities. A number of attempts have been made in the literature to relate this effect to the "minimum entropy production principle" of non-equilibrium thermodynamics. According to this principle, a dynamical system tunes itself up to the regime with minimum energy dissipation, corresponding to the minimum of $\dot{s}$ (supplied by $\delta\dot{s} = 0 \wedge \delta^2\dot{s} > 0$). For a frictional system, this is a regime with lowest coefficient of friction.

This effect can be viewed as erasing the information about the original roughness distribution of the surfaces in contact while the system is achieving an equilibrium stationary state of the lowest friction. Mortazavi and Nosonovsky [1] suggested Shannon entropy as a roughness quantitative parameter in order to characterize the evolution of a surface during the run-in stage toward a more ordered state (i.e., a state containing less information). The Shannon entropy of a rough surface is calculated by dividing the height (from the minimum to the maximum value) into $N$ bins and calculating the sum $-\sum_{n=1}^{N}(p_n/N)\log(p_n/N)$, where $p_n$ is a number of datapoints in *n*-th bin. Similarly to the analysys in the preceding section, the minimum corresponds to zero first derivative and positive second derivatives  Thus the stability criterion supplied by Eq, 12 can be modified as

$$S \equiv \min[-\sum_{n=1}^{N}(p_n/N)\log(p_n/N)] \tag{17}$$

## 2.4. Friction as an epiphenomenon combining nomothetic and idiographic

For most conventional textbooks of mechanics, dry friction is an external phenomenon,

which is postulated in the form of laws of friction (usually, the Coulomb-Amontons law) introduced in an arbitrary and ad hoc manner in addition to the constitutive laws of mechanics. Furthermore, the very compatibility of the Coulomb friction laws with the laws of mechanics is questionable due to the existence of the so-called frictional paradoxes or logical contradictions in the mechanical problems with friction. The Coulomb-Amontons law is not considered a fundamental law of nature, but an approximate empirical rule, whereas friction is perceived as a collective name for various unrelated effects of different nature and mechanisms, such as adhesion, fracture, and deformation, lacking any internal unity or universality.

Despite this artificial character of friction laws in mechanics, the Coulomb friction is a fundamental and universal phenomenon that is observed for all classes of materials and for loads ranging from nanonewtons in nanotribology to millions of tons in seismology. There is a contradiction between the generality and universality of friction and the artificial manner of how the friction laws are postulated in mechanics and physics. Remarkably, all diverse conditions and mechanisms of friction lead to the same (or at least similar) phenomena and phenomenological laws of friction.

If a thermodynamic approach is used consistently, the laws of friction and wear can be introduced in a much more consistent way. Historically, friction could be viewed as a fundamental force having the same status as the inertia force (nothing would move without inertia, nothing would stop without friction). The problem of the inertia force was one of the central issues of physics throughout the middle ages, until it was finally resolved by Galileo, leading to the foundation of modern mechanics by Newton in the late $17^{th}$ century. In contrast, the friction force was not usually seen as a fundamental force of nature and became somewhat marginal in modern mechanics.

The Galilean approach towards physics is in deconstruction of complex phenomena and finding their underling fundamental forces, thus building physics in a deductive (or nomothetic) manner. Those phenomena which do not fit the nomothetic scheme are considered idiographic. However, it has been suggested that the nomothetic-idiographic dichotomy is not appropriate for some situations. Friction represents an interesting example of a single emergent phenomenon which has several sources rather than a single source of origin.

While in principle it is expected that friction should be deduced from the 2nd law of thermodynamics, practically it is quite difficult to perform such a derivation. We have suggested in the past to use Onsager's linear formalism (by which linear viscous friction can be deduced) and an asymptotic transition from a 3D medium to a 2D interface for such a derivation [1].

At this point, there has been suggested no computational device using dry friction. However, there are demonstrations of computational potential of a related surface phenomenon – wetting of a solid by liquid. The possibility of a distributed computational device combining logical operations and chemical reactions are discussed in the folllowingf section.

## 3. Capillary phenomena: wetting, superhydrophopbicity, droplets, and bubbles

There are several other surface phenomena related to friction and adhesion, which are of interest for a logical and information analysis. Among them are adhesion, wetting, and the superhydrophobicity. The adhesion of water to a solid (called wetting) can be considered as a form of solid-liquid friction, and it has many features similar to the dry friction and finds its application in microfluidics [2, 14].

## 3.1. Superhydrophobic states and the Lotus effect

Superhydrophobicity is the surface roughness-induced non-wetting (the ability to repel water). The phenomenon mimics the Lotus leaf's ability to emerge clean from a dirty or muddy water, so it is often called the Lotus effect. Lotus is a symbol of purity in many Asian cultures. Thus the ancient Hindu poem Bhagavad Gita says about the seeker of truth "Having abandoned attachment, he acts untainted by evil, just as a lotus leaf is not wetted."

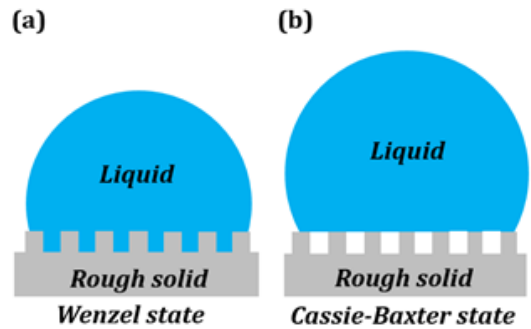

Figure 3. (a) A liquid droplet (a) in the Wenzel and (b) in Cassie-Baxter states.

When a water droplet is placed on a solid surface, the water surface forms a stable angle with the solid called the contact angle (CA) between 0° and 180°. Superhydrophobic surfaces have very high values of the CA>150°.

There are two states of a superhydrophobic surface: the Wenzel state (a homogeneous solid-water interface) and the Cassie state (composite solid-air-water interface with some air trapped between a water droplet and the solid within the cavities of the latter), Fig. 3 [15]. A transition between these two states is called a wetting transition and it has a number of remarkable properties.

Similarly to the friction with its "slip" and "stick" states, a logical binary predicate, M, can be introduced, so that M=true, when the surface is wetted (the Wenzel state) and the M=false when the surface is not wetted (the Cassie-Baxter state). Furthermore, there are situations with a "metastable" (very fragile and virtually unstable) Wenzel state. Therefore, much of the analysis suggested in the preceding sections for friction, can be applied to wetting.

*3.2. Droplet computers*

Prakash and Gershenfeld [16] have demonstrates that universal computation is possible in an all-fluidic two-phase microfluidic systems, i.e., with droplets or bubbles. Nonlinearity can be introduced into an otherwise linear, reversible, low–Reynolds number flow via bubble-to-bubble hydrodynamic interactions. The presence or absence of water can be defined using the binary predicate M=true (wetted state) and M=false (non-wetted state).

A bubble traveling in a channel represents a bit, providing the capability to simultaneously transport materials and perform logical control operations including AND/OR/NOT gates. These show the nonlinearity, gain, bistability, synchronization, cascadability, feedback, and programmability required for scalable universal computation. With increasing complexity in large-scale microfluidic processors, bubble logic provides an on-chip process control mechanism integrating chemistry and computation. It is not anticipated that bubble computational devices will substitute for traditional electronic ones, because the latter are much more powerful and faster. However, the bubble logic will find its application in the areas where small devices combine rapid chemical reactors and logical computation, such as lab-on-a-chip, micro/nanofluidic systems, and similar.

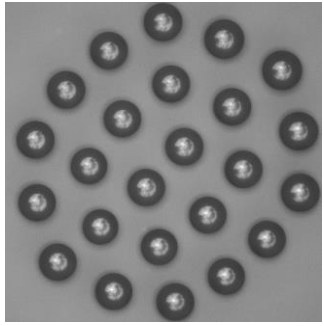

Figure 4. A levitating ordered cluster of 24 microdroplets [18].

Recent studies have also demonstrated that micro-bubbles can form ordered relatively stable structures such as levitating clusters of microdroplet (Fig. 4) with the number of droplets from one to hundreds [17, 18]. Such clusters provide new ways to manipulate droplets and potentially can lead to new computational applications. The Voronoi entropy, which is analogical to the Shannon entropy, is used to analyse information content of the clusters [17, 18].

*3.3. Logical gates for bubble and droplet logic*

The bubble logical gates were suggested by Prakash and Gershenfeld [16]. The OR and AND gate is shown in Fig. 5(a). Incoming bubbles can arrive through two inbound channels A and B, simultaneously or separately. There are also two outbound channels; however, these are not identical: one has a larger cross-section than the other. Consequently, when only one bubble arrives, it always goes through this thicker channel with lower pressure resistance, which therefore serves the "A ∨ B" output. When two bubbles arrive simultaneously, the second one will go through the thinner channel, which therefore serves the "A ∧ B" gate.

Similarly, the "¬A ∧ B" gate is realized by the setup shown in Fig. 5(b). The bubble incoming through the channel B will normally go through a thicker output channel with lower pressure. However, if simultaneously a bubble arrives through the channel A, it increases the pressure in the thicker channel above that in the thin one. Therefore, the bubble proceeds through the thinner channel, which now has lower pressure. Consequently, the thinner channel realizes the "A ∧ B" output, whereas the thicker channel realizes the "¬A ∧ B" output.

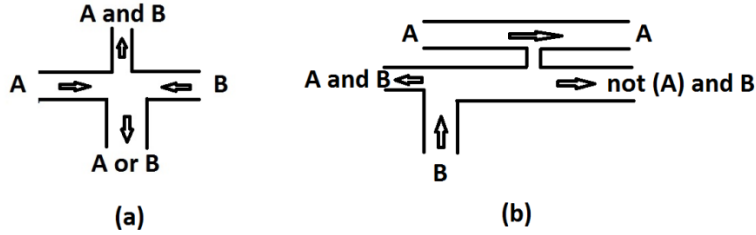

Figure 5. Bubble logical gates for (a) AND/OR and (b) NOT A AND B Boolean operations.

Logic implemented in low-Reynolds-number droplet hydrodynamics is asynchronous and thus prone to errors that prevent scaling up the complexity of logic operations. One particular limitation of pressure-based logic is that when many gates are attach, an interference can occur and channel's behaviour might depend on the state of neighbouring gates. To handle this parallelism-caused interference, non-classical logic may be needed, similar to that developed by Schumann [19].

A further development of droplet computational devices was suggested by Katsikis et al. [20], who proposed a method of algorithmic manipulation based on logic operations that automatically compute where droplets are stored or directed by enabling parallel control using synchronous universal logic. For that end, they employed a rotating magnetic field that enables parallel manipulation of arbitrary numbers of ferrofluid droplets on permalloy tracks. The coupling of magnetic and hydrodynamic forces between droplets enabled AND, OR, XOR, NOT and NAND logic gates and large-scale integration of droplet logic. While droplet devices are slower than electronic ones, the former still have a number of advantages. This include the possibility of slow processing with near-zero clock frequency, unlike electronic logic, where operations are generally performed at a fixed c lock frequency. Energy supply to all logic gates is by a 3D magnetic field, rather than a 2D network of conductive wires in conventional processors, which allows more efficient and precise energy supply [20].

The droplet and bubble logical and computational devices are conceptually similar to those build on the use of biological material, such as the conventional and unconventional reversible logic gates on *Physarum polycephalum* [21]. Given the advances of the droplet and bubble logic, some scientists even raised the question of "whether bubbles and droplets can think" [22]. Generally speaking, the new field of "digital microfluidics" has emerged from the integration of droplet/bubble logical and computational capabilities with the advances in microfluidics and lab-on-a-chip technology.

**4. Biomimetic vs. conventional design logic**

The Lotus-effect constitutes an example of a biomimetic approach in the surface science. The word "biomimetics" coined in the 1950s by the biophysicist Otto Schmitt was popularized in the 1970s. However, the idea of mimicking nature for artificial devices has existed since antiquity (for example, the idea of mimicking bird wings for men to fly in the myth of Icarus and the "robot"-like heroes such as the Golem) [2]. The reason biomimetics has become more and more popular since the 1970s is related to the increasing popularity of various "holistic" concepts.

During most of the 20$^{th}$ century, much technology was developed under the assumption that humans would eventually be able to transform nature in any way they wanted. Later humans understood that there is a lot to learn from nature. Traditional engineering approaches imply that you have an exact blueprint of your final product and an exact procedure how to make it. Living nature is very different. It has only general algorithms encoded in the DNA and hierarchical self-replicating structures, which adjust to the changing environment when they grow. So, instead of concentrating on a rigid homogeneous structure and on an exact solution to a problem, the biomimetic paradigm provides a much more flexible and holistic view.

At an even more fundamental philosophical level, one can view the transformative scientific revolution of the 17$^{th}$ century, which resulted in the development of the modern empirical scientific method, as the abandonment of Aristotle's approach towards nature as expressed in his physics and metaphysics. The new empirical method, as developed by Francis Bacon, Isaac Newton, Descartes, Leibnitz and other great minds of the 17$^{th}$ century, has led to the establishment of modern science, which resulted over the subsequent 300 years in an amazing number of discoveries and the transformative technologies of modernity. However, it became evident by the second half of the 20$^{th}$ century that such a "technocratic" approach also has its own limitations, leading to post-modern views on nature and humankind. The emergence of holistic approaches (including biomimetics) followed the same trend.

Biomimetics is increasingly popular in materials science as well as in the surface science where it yields new ways to control friction, adhesion and attachment/detachment [2]. The most successful examples of biomimetic application in surface science are the Lotus-effect, the gecko-effect and the shark-skin effect. The gecko-effect is the strong and controlled adhesion due to a special hierarchical structure of the gecko foot. The shark-skin effect is the ability to reduce drag resistance in water flow due to special orientation of micro-riblets. Dozens or even hundreds of other potential biomimetic approaches are discussed in the literature, including bio-inspired logical and game theory applications as those considered by Schumann and co-workers [23]. This brings the need to consider biomimetic design logic seriously, even although the latter might not always constitute "logic" in a technical sense of the word.

**5. Conclusions**

There is a number of important logical and informational implications for friction and wetting. Combining dry friction with the rest of mechanics involves logical paradoxes, the so-called Painlevé paradoxes. These paradoxes correspond to a broader phenomenon of frictional instabilities, vibrations, and self-organization. Information (or entropy) production constitute an important measure of frictional self-organization. Friction is similar to the process of erasing the information. Furthermore, friction is an example of a process with is not deduced from a single underlying effect but instead from a number of different mechanisms. Friction is related to a broader class of surface effects including adhesion, wetting and the superhydrophobicity (the biomimetic Lotus effect). These include a logical droplet computing for microfluidic and lab-on-chip applications as well as general biomimetic design considerations. Therefore, surface phenomena, such as friction and wetting, can be used for unconventional distributed logical and computational devices combining logical operations and chemical micro-reactors in one device.

**Acknowledgement**. Supported in part at the Technion by a fellowship from the Lady Davis Foundation.

*the 8th International Conference on Bioinspired Information and Communications Technologies*, 9-16